\documentclass[prd,aps,showpacs,nofootinbib,superscriptaddress,preprint]{revtex4-1}%
\usepackage{graphicx,color,amsmath,amsxtra}
\usepackage{epsf}
\usepackage{amssymb}
\usepackage{enumerate}
\usepackage{hhline}
\usepackage{array}
\usepackage{tabularx}
\usepackage{slashed}
\usepackage{bm}
\usepackage{multirow}

\newcommand{\ba}{\begin{array}}
\newcommand{\ea}{\end{array}}

\newcommand{\eq}[1]{(\ref{#1})}

\newcommand{\newc}{\newcommand}
\newc{\be}{\begin{equation}}
\newc{\ee}{\end{equation}}
\newc{\beq}{\begin{eqnarray}}
\newc{\eeq}{\end{eqnarray}}
\newc{\bes}{\begin{subequations}}
\newc{\ees}{\end{subequations}}

\begin{document}

\title{K-theoretic classification of fermionic operator mixings in holographic conformal field theories}

\author{Shih-Hao Ho}

\author{Feng-Li Lin\footnote{linfengli@phy.ntnu.edu.tw}}
\affiliation{Department of Physics, National Taiwan Normal University, Taipei 116, Taiwan}

%\date{\today}

%%%%%%%%%%%%%%%%%%%%%
%  Abstract
%%%%%%%%%%%%%%%%%%%%%
\begin{abstract}
   In this paper, we apply the K-theory scheme of classifying the topological insulators/superconductors to classify the topological classes of the massive multi-flavor fermions in anti-de Sitter (AdS) space. In the context of AdS/CFT correspondence, the multi-flavor fermionic mass matrix is dual to the pattern of operator mixing in the boundary conformal field theory (CFT). Thus, our results classify the possible patterns of  operator mixings among fermionic operators in the holographic CFT.

\end{abstract}
%%%%%%%%%%%%%%%%%%%%%

\maketitle

%$\tableofcontents

\section{Introduction}

In the past few years, the discovery of the topological insulators/superconductors (TI/TSc) \cite{HasanKane,Qi2011} shed a new light on our understanding about the simple free relativistic fermion systems, which manifest as the corresponding low energy theory and are otherwise thought to be trivial in the usual consideration of condensed matter physics.  The topological characteristics of these system are guaranteed by some global symmetries such as time reversal ($T$), charge conjugation ($C$) or parity ($P$) \cite{Altland, LudwigRyu}. hence are dubbed as the symmetry-protected topological oder (SPT) \cite{spt}. Moreover, by exploiting the Clifford algebra of the Dirac fermion systems one can put the above classification scheme into the framework of K theory analysis \cite{Kitaev,Wen1}. The K-theory analysis yields the classifying space of the possible mass deformations, which is then constrained by the Clifford algebra formed by  the symmetry generators  in the Gamma matrix representation.  In Ref. \cite{Ho:2012gz}, we have applied the scheme to the problems in the context of high energy physics and resolve some issue \cite{Jackiw:1975fn,McGreevy:2011if} on the possible number of Majorana modes in the presence of $SU(2)$ Witten anomaly \cite{Witten:1982fp}.

In this paper, we would like to continue our application of the K-theoretic classification scheme to the problems in the context of high energy physics. This time we will classify the massive multi-flavor free fermions in the anti-de Sitter (AdS) space. Our motivation is two-fold. The first is to follow the pioneer works \cite{Ryu:2010hc1,Ryu:2010hc2} to find out more possible realizations of the topological ordered systems in the context of string theory.  In contrast to the Minkowski space, there is a confined potential in the AdS space so that its asymptotic boundary can be thought as a co-dimensional one defect which can then host the non-normalizable fermionic zero modes.  In this way, the bulk classification can also yield the classification of the edge modes as in the usual context of bulk-edge correspondence \cite{Jackiw:1975fn,bulk-edge,Ho:2012gz}.

The second motivation is to see the implication of the K-theoretic classification scheme in the context of AdS/CFT correspondence \cite{Maldacena:1997re1,Maldacena:1997re2,GKPW1,GKPW2}. Unlike the identification of the anyonic excitations in the dual CFT as done in Refs. \cite{Hartnoll:2006zb,Kawamoto:2009sn}, here it is related to issues of operator mixings and level crossings \cite{Korchemsky:2015cyx}.

In AdS/CFT, it is rare to consider the multi-flavor fields as they will be holographic dual to a set of operators. However, it was known some time ago \cite{Constable:2002vq} that the bulk quantum corrections will in general induce mass mixings of the bulk fields, which thus  are then dual to the operator mixings in the dual CFT as the masses of the bulk fields are related to the conformal dimensions of the dual operators.  In this context, our massive multi-flavor bulk fermion systems can be thought as effective field theory taking care of bulk quantum corrections and the resulting mass mixings. This can then be dual to the fermionic operator mixings due to the sub-leading corrections beyond the large $N$ limit. Thus, the K-theoretic classification of the fermion mass matrix is dual to the classification of distinct patterns of fermionic operator mixings. One can then imagine that there are possible level crossings as one tune $N$ or some relevant parameters such that the operator mixing pattern changes from one to the other.

Though it is usually technically hard to obtain the explicit form of the bulk quantum corrections and the resulting mass mixings, it is still sensible to ask what are the possible patterns of dual operator mixings by requiring some global symmetries such as $C$, $P$ or $T$. This is similar to the philosophy adopted in the study of topological order for which the classifications are usually done without reference to explicit solvable models.

Our paper is organized as follows. In the next section we will outline the setup for our consideration, namely, the AdS space as a topological insulator with its asymptotic boundary hosting the fermionic edge modes.  In Sect. \ref{sec3} we pedagogically review the basics of the K-theory analysis in classifying the TI/TSc. Especially we formulate the scheme in the language of high energy physics so that we can directly apply it in the context of high energy physics as well as AdS/CFT in the following next two sections.  In Sect. \ref{sec4} we apply the results in Sect. \ref{sec3}  to classify the topological phases of the massive free real fermions in high energy physics.  We then use these results with some other constraints to classify the symmetry-protected topological orders for the holographic CFTs in Sect. \ref{sec5}. Our main results are summarized in Table \ref{tableHEP} and \ref{tableF}. Finally, we conclude our paper in Sect. \ref{sec6} with some discussions.

\section{Anti-de Sitter space as TI/TSc}\label{sec2}

\begin{center}
\begin{figure}[bbp]
\begin{tabular}{ll} \nonumber
\begin{minipage}{80mm}
\begin{center}
\unitlength=1mm
\resizebox{!}{5cm}{
\includegraphics[scale=0.35]{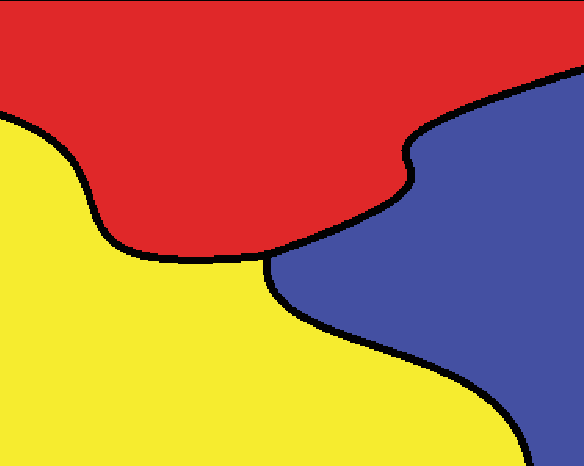}  }
\end{center}
\end{minipage}
\end{tabular}
\caption{(Color online) A sketch of phase diagram of the gapped systems in the configuration (parameter) space. The regions with different colors correspond to different topological phases, and are separated by the phase boundaries which host the gapless excitations living on the topological defects }
\label{phase diagram}
\end{figure}
\end{center}

The topological phases are not characterized by order parameters such that they cannot be classified by the usual Landau-Ginzburg paradigm in the context of spontaneous symmetry breaking. Instead the topological phases are characterized by their robustness against different types of perturbations. For the gapped systems the robustness are defined by none of gap closing (level crossing) as one tunes the relevant parameters or perturbations. Therefore, two different topological phases are always separated by a phase boundary in the configuration space of some physical parameters, on which the gapless modes appear (Fig. \ref{phase diagram}).  The free massive multi-flavor fermions are the simplest gapped systems for such a consideration. The K-theory analysis for these systems is figure out the topological properties of the parameter/configuration space \cite{Kitaev,Wen1}, i.e., an index theorem by counting the gapless modes located on the phase boundary according to the nontrivial homotopy group of the parameter/configuration space.

In this work, we like to apply the K-theoretic classification scheme developed in Refs. \cite{Kitaev,Wen1} to the multi-fermions in $(d+1)$-dimensional AdS space. Follow the original setup of Jackiw-Rebbi model \cite{Jackiw:1975fn,Ho:2012gz}, we will consider the massive free fermions a scalar condensate $\varPhi$ as following:
\beq \label{A01}
S_f=  \int {\rm d}^{d+1} x \sqrt{-g} \;   {\rm i}  (\bar{\psi}  \slashed{D}  \psi - \varLambda \bar{\psi} \varPhi  \psi - m \bar{\psi} \psi ) .
\eeq
Above we have omitted the gauge and flavor indices of fermion $\psi$ and  flavor indices of matrices $\varLambda$ and $m$.

In the context of AdS/CFT correspondence, a bulk field is dual to a CFT operator. For example, the bulk metric is dual to the CFT's stress tensor, and a scalar field of mass $m$ is dual a scalar operator of conformal dimension $\varDelta={d\over 2} \pm  m$. The relation between $m$ and $\varDelta$ is obtained by the GKP/W prescription of evaluating the CFT correlator  $1/|x-y|^{2\varDelta}$ from the corresponding bulk on-shell action \cite{GKPW1,GKPW2}.  Similarly, a fermionic bulk field is dual to a fermionic CFT operator with the same relation between $m$ and $\varDelta$.

Besides, for considering multi-fields in the bulk, one can also introduce some interactions among them, which will then introduce higher point functions and operator mixing in the dual CFT. In this paper, we will consider two types of bulk ``interactions". The first type is to introduce the Yukawa coupling between bulk fermions and scalars respecting the $SU(2)$ symmetry. This kind of terms is to enrich the topological structures in analogy to the classification of TI/TSc. In the context of AdS/CFT correspondence, this kind of $SU(2)$ symmetry is naturally to be realized as part of the $R$-symmetry or flavor symmetry of the dual CFT.  The second type is to introduce a generic mass matrix for the multi-fermion fields, this will be dual to the operator mixings in the CFT as the mass eigenvalues are related to the conformal dimensions of CFT operators. Our goal of this paper is to classify the total fermion mass matrix, i.e., the free fermion mass matrix plus the Higgs mass terms from the Yukawa coupling, in the K-theoretic framework.

From Eq. \eq{A01}, we can obtain the Dirac equation in the AdS space as following
\be \label{A03-4}
  h \psi=\omega \psi,
\ee
where the Hamiltonian
\be\label{effH}
h=   {\rm i}  \;  \varGamma^0 [\; r^2 \varGamma^r \partial_r - r( \varGamma^a \partial_a + m + \varLambda \varPhi) ]+ {\rm i} {d\over 2} r \varGamma^{0}\varGamma^r,\;
\ee
with respect to the following metric for $AdS_{d+1}$ space,
\be
{\rm d}s^2=r^2 (-{\rm d}t^2 + \delta_{ab} {\rm d}x^a {\rm d}x^b)+{{\rm d}r^2\over r^2}.
\ee
  Note that the derivation of Eqs. \eq{A03-4}  and \eq{effH} can be found in Ref. \cite{Iqbal:2009fd}.

In comparison with Eq. \eq{effH} for AdS case, the Hamiltonian for the massive free fermions in $(d+1)$-dimensional Minkowski space is
\be\label{effH0}
h=  {\rm i}  \varGamma^0( \varGamma^i \partial_i - m - \varLambda \varPhi) :={\rm i} (\alpha^i \partial_i -\bm{M}),
\ee
where $i=1,2,\cdots,d$ and $\alpha^i:=\varGamma^0\varGamma^i$. The total mass matrix $\bm{M}$ is defined by
\be \label{mass M}
\bm{M}=\varGamma^0\otimes (m +\varLambda \varPhi).
\ee

\begin{center}
\begin{figure}[bbp]
\begin{tabular}{ll} \nonumber
\begin{minipage}{120mm}
\begin{center}
\unitlength=1mm
\resizebox{!}{6cm}{
\includegraphics[scale=1]{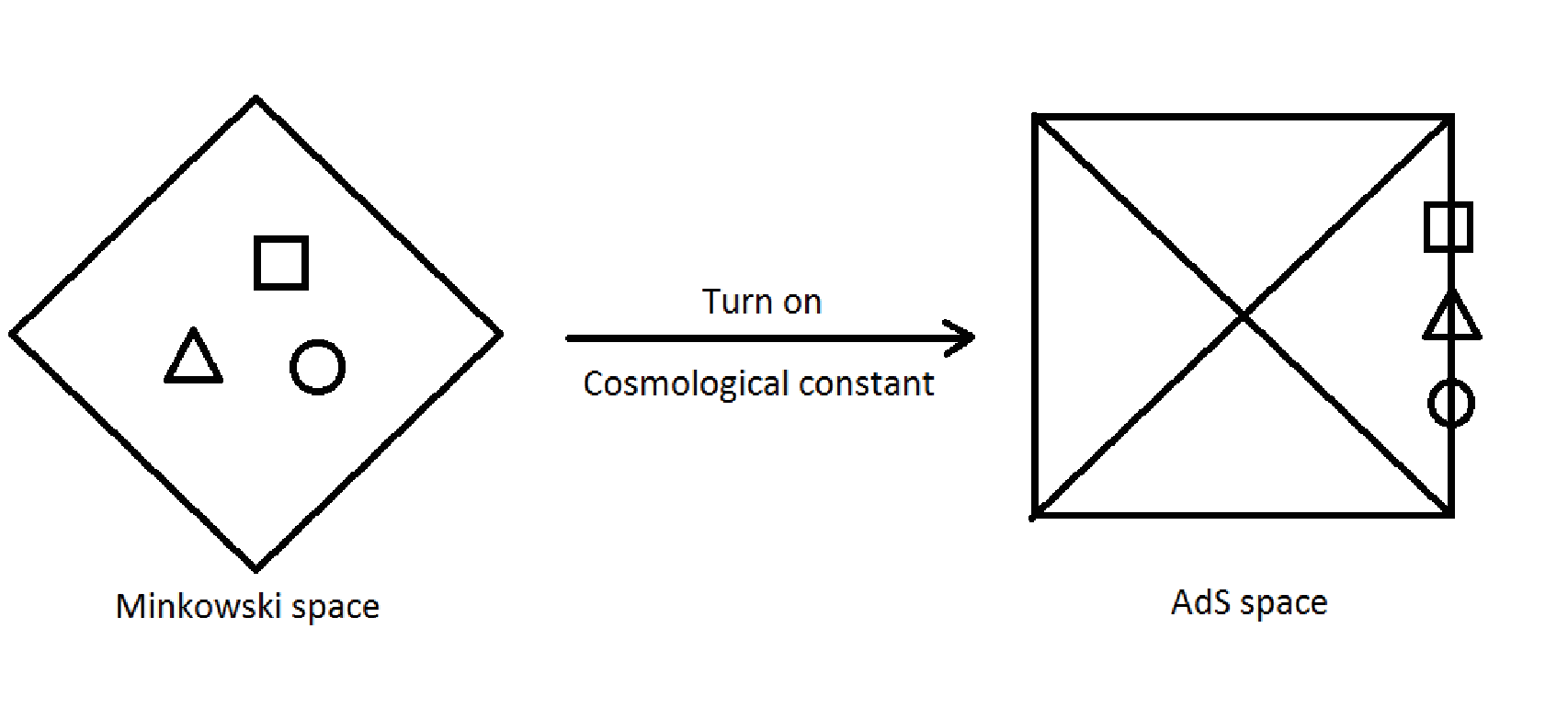}  }
\end{center}
\end{minipage}
\end{tabular}
\caption{A sketch of the idea of viewing the Minkowski and AdS spaces as topological insulators related by turning on the cosmological constant. The symbols in the figure are some topological excitations  }
\label{cartoon}
\end{figure}
\end{center}

The difference between Eqs. \eq{effH} and \eq{effH0} is the appearance of gravitational potential in Eq. \eq{effH} due to the cosmological constant.  The gravitational potential is a confining one and its effect is to close the mass gap at the AdS asymptotic boundary. This can be also understood from the fact that the on-shell Hamiltonian at the asymptotic boundary is also the Hamiltonian of the dual CFT, which by definition is gapless, i.e., zero Hamiltonian. Therefore, the AdS space can be thought as a TI/TSc with a asymptotic boundary hosting the gapless edge modes, see Fig. \ref{cartoon}, i.e., a kind of bulk-edge correspondence of TI/TSc realized by the gravitational potential.  See also Ref. \cite{Kaplan:2011vz} for similar proposal in the different context.

Besides, we should emphasize that the bulk/edge correspondence mentioned here should not be confused with the AdS/CFT correspondence. We just use the former to classify the bulk mass patterns, and through the latter this classification can be translated to the one for the patterns of the operator mixings in the dual CFT.  Also note that, though the CFT is gapless, the dual bulk theory is gapped as we have introduced the fermion mass terms.  Moreover, the AdS boundary is treated as a defect for the bulk gapped theory, which can localize the topological modes. However, these edge modes are gapless so that we did not introduce gapped quantities in the AdS boundary where the dual CFT lives.

Mathematically, the topological classes of bulk modes of a TI/TSc are classified by the homotopy group  $\pi_0(\mathcal{M})$ which counts the number of simply connected components of the parameter/configuration space $\mathcal{M}$. On the other hand, the topological classes of the corresponding edge modes on a defect of spatial dimension is given by the homotopy group $\pi_{d-d_b-1}(\mathcal{M})$.  This is one version of the mathematical statement for the bulk-edge correspondence \cite{Wen1}. Note that if the defect is of co-dimensional one, then both the bulk and edge modes are classified by the same homotopy group $\pi_0(\mathcal{M})$.

As alluded to in the introduction, the fermion mass matrix encodes the patterns of the operator mixings of the dual CFT in the context of AdS/CFT correspondence. Thus, the classification of the mass matrix $\bm{M}$ of Eq. \eq{mass M} is dual to the classification of the patterns of the operator mixings. As we tune the relevant parameters so that the topological class of the mass matrix changes, one will then expect the corresponding level crossings happen in the dual CFT side.  The remaining part of this paper will be devoted to the details of K-theoretic classification of the mass matrix $\bm{M}$.

Finally, what we have done is the same as classification of the SPT phases of the free fermion systems. Therefore, we will also sometimes refers the ``classification of the patterns of fermionic operator mixings" as the ``classification of the fermion SPT phases" in a loose way.

\bigskip
\bigskip

\section{K-theory classification scheme}\label{sec3}

In this section we will review the basics of the K-theory analysis in classifying the topological phases of the massive free fermions given in Refs. \cite{Kitaev,Wen1}. Especially, we try to pedagogically reformulate the scheme in the language of high energy physics, to which our setting belongs.

The Hamiltonian of the massive free fermion system is given in Eq. \eq{effH0}.  Note that in the context of high energy physics, the spacetime part of $\alpha^i$ and $\bm{M}$ are fixed by the Lorentz symmetry.  Furthermore,  we need to require the hamiltonian is hermitian for the physical systems, i.e., $h^{\dagger}=h$, this then yields
\be\label{hermitian-aM}
(\alpha^i)^{\dagger}=\alpha^i,\qquad \bm{M}^{\dagger}=-\bm{M}.
\ee
Note that by the definition of $\alpha^i$ the first equation is automatically satisfied, however, the second equation will constrain the gauge and flavor structures of $\bm{M}$, i.e.,
\be\label{hermitian-mL}
(m+\varLambda \varPhi)^{\dagger}=m+\varLambda\varPhi.
\ee

The basic ideas of the K-theory analysis is to classify the configuration (or parameter) space of the mass $\bm{M}$ which forms some  Clifford algebra with some symmetry generators. The types of the Clifford algebras will then determine the possible topological phases for the free massive fermions. Among these symmetry generators, the most important ones are the charge conjugation ($C$) and time reversal reversal ($T$) symmetry operations defined by
\bes\label{CTsym}
\beq\label{CTsym0-C}
C \psi(t,\vec{x})C^{-1}=U_c \psi^*(t,\vec{x}),
\\
T\psi(t,\vec{x})T^{-1}=U_t \psi(-t,\vec{x}),
\eeq
\ees
where $U_c$ and $U_t$ are unitary matrix representations of $C$ and $T$, respectively.

Usually, in the textbook the $U_c$ and $U_t$ are taken some specific forms after fixing the convention for the representation of Lorentz symmetry, for example see Ref. \cite{PeskinQFT}. However, in general the choices of $U_c$ and $U_t$ are not unique, particularly when there are some low energy background such as the uniform scalar condensates. Specifically, the explicit forms of $U_c$ and $U_t$ depends on the scalar and fermion representations of a given group $G$. To some extent, turning on the different scalar condensates correspond to tuning the vacuum properties.  On top of the chosen vacuum, we could have different topological fermionic phases.

 \subsection{Flat-band condition}

We will adopt K-theory analysis to classify the gapped topological phases, this means that  we are only interested in the $h$ with nonzero eigenvalues, i.e., $\det h\ne 0$. Moreover, as only the topological property of such $h$ is concerned, we just need to know the numbers of positive and negative energy eigenvalues, not its detailed spectroscopy. For this purpose we will impose the so-called ``flat-band" condition
\be
h^2=(\vec{p} \cdot \vec{p}+1)\otimes \mathbf{1} ,
\ee
where $\vec{p}$ is the spatial momentum and $\mathbf{1}$ is the unit matrix in the gauge and flavor spaces. This then gives
\bes \label{flatband}
\beq
\label{flatband-1}
\{\alpha^i \otimes \mathbf{1} ,\alpha^j \otimes \mathbf{1} \}&& =2\delta^{ij} \otimes \mathbf{1}, \\
\label{flatband-2}
\{\alpha^i,\bm{M}\} && =0,\\
\label{flatband-3}
\bm{M}^2 && =-1 \otimes \mathbf{1}.
\eeq
\ees
In obtaining Eq. \eq{flatband-2}, we have assumed $\bm{M}$ is uniform. Sometimes, we will not distinguish $\alpha^i$ and $\alpha^i \otimes \mathbf{1}$.

The Clifford algebra \eq{flatband}  will be the starting point for the K-theory analysis in classifying the fermionic gapped phases.

 \subsection{K-theory classification for complex fermions}

We first consider the cases without the charge conjugation and time reversal symmetries. In these cases, Eqs. \eq{hermitian-aM} and \eq{flatband}  are all what we need.  In the context of high energy physics \footnote{The key difference in adopting the above scheme for the high energy and condensed matters is that in the former $\alpha^i$'s do not mix with the gauge and flavor structure. On the other hand, in the condensed matter \eq{effH0} is the low energy effective model so that $\alpha^i$ may be mixed with other non-spacetime structures. }, $\alpha^i$ are the space-time Gamma matrices fixed by the Lorentz symmetry, thus  Eq. \eq{flatband-1} is by definition automatically satisfied, and   Eqs. \eq{flatband-2} and \eq{flatband-3} constrain the possible mass matrix $\bm{M}$.   Note that Eq. \eq{flatband} form a Clifford algebra $Cl(d,1)$, and given $\alpha^i$'s, the configuration space $\mathcal{M}$ will be determined by $Cl(d,1)$ for complex matrices, and denoted by $C_{d}$. The space $C_p$ and its homotopy group is given in Table \ref{tableC} \cite{Kitaev,Wen1}.  This then classifies the two possible topological phases for the free complex fermions with $\mathcal{M}=C_d$.

Note that the nontrivial $\pi_0(C_{p={\rm even}})=\bm{Z}$ means that there are nontrivial topological fermionic phases which can be labelled by some integer topological invariants.

\begin{table} %[htdp]
\caption{The space $C_p$ and its homotopy group $\pi_0(C_p)$.}
\begin{center}
\begin{tabular}{ c c c c c}
\hline
$p$ mod 2  & $C_p$ & $\pi_0(C_p)$ & $\pi_1(C_p)$ & $\pi_2(C_p)$\\
\hline
0 & ${U(l+m) \over U(l) \times U(m)}\times \bm{Z}$  & $\bm{Z}$ & 0 &$\bm{Z}$\\
1 & $U(n)$  & 0& $\bm{Z}$ & 0  \\
\hline
\end{tabular}
\end{center}
\label{tableC}
\end{table}%

We can also classify the boundary excitations living on the defects by $\pi_{d-d_b-1}(C_d)$ where $d_b$ is the spatial dimension of the defect.  Using the Bott periodicity theorem
\be\label{BottC}
\pi_n(C_p)=\pi_0(C_{p+n}),
\ee
 we obtain $\pi_{d-d_b-1}(C_d)=\pi_0(C_{d_b+1})$, which is $d$-independent. It says that point defects host no fermionic excitations but the line defects host fermionic excitations classified by integer labels.

\subsection{Charge conjugation and time reversal symmetries}

Now we move to the cases with the charge conjugation and time-reversal symmetries defined in Eq. \eq{CTsym}.  To require these symmetries \footnote{We assume our theory is $CPT$ invariant so that $P$ symmetry will be determined by $C$ and $T$.} implies the following conditions on the Hamiltonian \cite{PeskinQFT}
\bes\label{CTsym-h}
\beq
\label{CTsym-h-C}
h(\vec{\partial},\vec{x})&=&-[U_c^{\dagger} h(-\vec{\partial},\vec{x}) U_c]^{\rm T}, \\
\label{CTsym-h-T}
h(\vec{\partial},\vec{x})&=&U^{\dagger}_t h^*(\vec{\partial},\vec{x}) U_t.
\eeq
\ees
These then yield the conditions on $\alpha^i$'s and $\bm{M}$ as follows:
\bes\label{CTsym-aM}
\beq
\label{CTsym-a-C}
\alpha^i U_c &=& U_c (\alpha^i)^*, \\
\label{CTsym-M-C}
 \bm{M} U_c &=& U_c \bm{M}^*, \\
\label{CTsym-a-T}
 (\alpha^i)^* U_t &=&-U_t \alpha^i, \\
\label{CTsym-M-T}
\bm{M}^* U_t &=&-U_t \bm{M}.
\eeq
\ees
Using Eq. \eq{hermitian-aM}, we have replaced $(\alpha^i)^{\rm T}$ by $(\alpha^i)^*$ in Eq. \eq{CTsym-a-C} and $\bm{M}^{\rm T}$  by $-\bm{M}^*$ in Eq. \eq{CTsym-M-C}, respectively.

Moreover, the charge conjugation and time reversal are $\bm{Z}_2$ symmetries,  they are characterized by the following $\bm{Z}_2$ parameters defined in the relations
\bes \label{G01}
\beq
\label{G01-1} && U_c U_c^*=s_C, \\
\label{G01-2} && U_t U_t^*=s_T, \\
\label{G01-3} && U_t U_c =s_{TC} U_c^* U_t^*,
\eeq
\ees
where the parameters $s_C$, $s_T$ and $s_{TC}$ are taking values of $\pm 1$.

\subsection{K-theory classification for real fermions}

To sort out the connectedness of the configuration space $\mathcal{M}$ we need to figure out what is the Clifford algebra formed by Eqs. \eq{flatband}, \eq{CTsym-aM} and \eq{G01}.  However, this is not possible because Eqs. \eq{CTsym-aM} and \eq{G01} involve complex conjugations, unless for the real representations, i.e., $\alpha^i$, $\bm{M}$, $U_c$ and $U_t$ are all real. In the discussions of this section we will simply assume there exists such a real representation, and in the later sections we will examine the existence condition case by case. Especially, in such a representation, $M$ is real and anti-symmetric and $\alpha^i$'s are real and symmetric such that the hamiltonian $h$ is pure imaginary and anti-symmetric, a condition required by Fermi statistics. Under such a real representation, the relations \eq{CTsym-aM} and \eq{G01} become
\be\label{commuCT}
[U_c,M]=[U_c,\alpha^i]=0, \qquad  \{U_t,\alpha^i\}=\{U_t,\bm{M}\}=0
\ee
and
\be\label{ParityCT}
U_c^2=s_C, \qquad U_t^2=s_T \qquad \mbox{and} \; U_t U_c=s_{TC} U_c U_t.
\ee

From Eqs. \eq{flatband}, \eq{commuCT} and \eq{ParityCT},  it is easy to see that $U_t$, $\alpha^i$ and $\bm{M}$ form a Clifford algebra, and $U_c$ may join to form a bigger Clifford algebra depending on the parity parameters $s_C$, $s_T$ and $s_{TC}$.  Given the $\alpha^i$, $U_c$ and $U_t$, the configuration space of possible real and anti-symmetric matrices $\bm{M}$ will be determined and denoted by $\mathcal{M}^{s_{TC}}_{s_T,s_C}$.   In practical, $\mathcal{M}^{s_{TC}}_{s_T,s_C}$ is determined by considering all the possible rotations which change only $\bm{M}$ but not the other generators in the Clifford algebra.

To determine $\mathcal{M}^{s_{TC}}_{s_T,s_C}$ we should consider them case by case.
\begin{itemize}

 \item If $s_{TC}=1$ and $s_C=1$, then $U_c$ commutes with all other operators so it can be set to one and will not contribute to the resultant Clifford algebra. Then,
 \begin{itemize}
 \item [(1)] If $s_T=1$, the resultant Clifford algebra is $Cl(d+1,1)$. Given all other operators except $\bm{M}$, the configuration space $\mathcal{M}^{+}_{++}$ is denoted by $R^0_{d+1}$.

 \item [(2)] Similarly, for $s_T=-1$, the resultant Clifford algebra is $Cl(d,2)$ and $\mathcal{M}^{+}_{-+}=R^1_d$.

 \end{itemize}

 \item If $s_{TC}=1$ but $s_C=-1$, then $U_c$ still commutes with all other operators but square to $-1$. Naively, we would set $U_c={\rm i} \mathbf{1}$ but it is not real.  Because we choose $U_c$ to be real, we should instead set $U_c=\mathbf{\epsilon} \otimes \mathbf{1}$, where $\epsilon$ is the epsilon tensor, i.e., ${\rm i}\sigma^y$. In such a case,  we can set the real anti-symmetric $\bm{M}=\mathbf{1} \otimes H_{\rm a} +\mathbf{\epsilon} \otimes H_{\rm s}$ where $H_{\rm s}$ and $H_{\rm a}$ are real symmetric and real anti-symmetric, respectively. Or it can collapse to $H=-H_{\rm s}+{\rm i} H_{\rm a}$.  We also collapse real symmetric matrix $\alpha^i=\mathbf{1}\otimes A^i_{\rm s}+\mathbf{\epsilon} \otimes A^i_{\rm a}$ into complex ones $A^i=A_{\rm s}^i+{\rm i} A_{\rm a}^i$ in the similar manner. Then Eq. \eq{flatband} implies
\bes \label{flatbandC}
\beq
\label{flatbandC-1}
\{A^i,A^j\}&& =2\delta^{ij}, \\
\label{flatbandC-2}
\{A^i,H\} && =0,\\
\label{flatbandC-3}
H^2 && =\mathbf{1}.
\eeq
\ees
Besides, we should also take time reversal $U_t$ and the associated \eq{CTsym-a-T} and \eq{CTsym-M-T} into account. Then,
 \begin{itemize}
 \item [(3)] For $s_T=1$, we can decompose $U_t=\mathbf{1} \otimes T_{\rm s} + \mathbf{\epsilon} \otimes T_{\rm a}$ and collapse it into $T=T_{\rm s}+{\rm i} T_{\rm a}$, where $T_{\rm s}$ and $T_{\rm a}$ are real symmetric and anti-symmetric, respectively.  So, Eqs. \eq{CTsym-a-T} and \eq{CTsym-M-T} imply that
\bes \label{TaMC1}
\beq
\label{TaMC1-1}
\{T,A^i\}&& =0, \\
\label{TaMC1-2}
\{T,H\} && =0,\\
\label{TaMC1-3}
T^2 && =\mathbf{1}.
\eeq
\ees
 Combining Eqs. \eq{flatbandC} and \eq{TaMC1} yields the Clifford algebra $Cl(d+2,0)$ for the complex matrices, and the resultant $\mathcal{M}^{+}_{+-}=C_{d+1}$.

 \item [(4)] Similarly for $s_T=-1$, we can collapse $U_t=\mathbf{1} \otimes T_{\rm a} + \mathbf{\epsilon} \otimes T_{\rm s}$ into $T=-T_{\rm s} + {\rm i} T_{\rm a}$ so that Eqs. \eq{CTsym-a-T} and \eq{CTsym-M-T} again imply Eq. \eq{TaMC1}. Thus, again $\mathcal{M}^{+}_{--}=C_{d+1}$.

\end{itemize}

Note that the configuration space $\mathcal{M}$ in such cases is $C_{d+1}$ which are different from $C_d$ for the cases without charge conjugation and time reversal symmetries.

 \item If $s_{TC}=-1$, then all the operators are anti-commuting with each other except $[U_c,\bm{M}]=0$. However, $U_tU_c$ anti-commutes with all other operators, including $\bm{M}$. Thus, $\alpha^i$'s, $U_t$, $U_tU_c$ and $\bm{M}$ form a Clifford algebra, which is characterized by $s_T=U_t^2$ and $s_{TC}s_Ts_C=(U_tU_c)^2$.  We then have
  \begin{itemize}
  \item [(5)] For $s_T=1$ and $s_C=1$, the Clifford algebra is $Cl(d+1,2)$ and $\mathcal{M}^-_{++}=R^1_{d+1}$.

  \item [(6)] For $s_T=-1$ and $s_C=1$, the Clifford algebra is $Cl(d+1,2)$ and $\mathcal{M}^-_{-+}=R^1_{d+1}$.

  \item [(7)] For $s_T=1$ and $s_C=-1$, the Clifford algebra is $Cl(d+2,1)$ and $\mathcal{M}^-_{+-}=R^0_{d+2}$.

  \item [(8)] For $s_T=-1$ and $s_C=-1$, the Clifford algebra is $Cl(d,3)$ and $\mathcal{M}^-_{--}=R^2_d$.

 \end{itemize}

 \end{itemize}

Moreover, the Bott periodicity of the Clifford algebras gives  \cite{Kitaev, Wen1}
\bes\label{BottR}
\beq \label{BottR-1}
R^q_p&=&R^0_{p-q \; {\rm mod} \; 8},\\
 \pi_n(R_p^0)&=&\pi_0(R^0_{p-n}). \label{BottR-2}
\eeq
\ees
We summarize the above classification in Table \ref{tableM}.
\begin{table}%[htdp]
\caption{The classifying spaces $\mathcal{M}^{s_{TC}}_{s_T s_C}$.}
\begin{center}
\begin{tabular}{ c c c c c c c c c }
\hline
$\mathcal{M}^{s_{TC}}_{s_T s_C}$ & $\mathcal{M}^+_{++}$  & $\mathcal{M}^+_{-+}$ & $\mathcal{M}^+_{+-}$ & $\mathcal{M}^+_{--}$ & $\mathcal{M}^-_{++}$ & $\mathcal{M}^-_{-+}$ & $\mathcal{M}^-_{+-}$ & $\mathcal{M}^-_{--}$
\\
\hline
Bulk phase & $R^0_{d+1}$ & $R^0_{d-1}$ &  $C_{d+1}$ &  $C_{d+1}$ & $R^0_d$ & $R^0_d$ & $R^0_{d+2}$ & $R^0_{d-2}$
\\
Defects & $R^0_{d_b+2}$ & $R^0_{d_b}$ &  $C_{d_b}$ &  $C_{d_b}$ & $R^0_{d_b+1}$ & $R^0_{d_b+1}$ & $R^0_{d+3}$ & $R^0_{d_b-1}$
\\
\hline
\end{tabular}
\end{center}
\label{tableM}
\end{table}%

\begin{table}%[htdp]
\caption{The space $R_p^0$ and its homotopy group $\pi_0(R^0_p)$.}
\begin{center}
\begin{tabular}{c c c}
\hline
$p$ mod 8  &$R^0_p$  & $\pi_0(R^0_p)$   \\\hline
0 & ${O(2n)\over U(n)}$ &$\bm{Z}_2$  \\
1& $O(n)$ & $\bm{Z}_2$\\
2&  ${O(l+m)\over O(l) \times O(m)} \times \bm{Z}$ & $\bm{Z}$\\
3 &  ${U(n) \over O(n)}$ &0 \\
4 & ${Sp(n) \over U(n)}$  &0\\
5 & $Sp(n)$ &0\\
6 &  ${Sp(l+m)\over Sp(l) \times Sp(m)} \times \bm{Z}$ &  $\bm{Z}$\\
7&${U(2n) \over Sp(n)}$&0\\
\hline
\end{tabular}
\end{center}
\label{tableR}
\end{table}%

Using these relations and the homotopy group of $R^0_p$ given in Table \ref{tableR} \cite{Kitaev,Wen1}, we can classify the topological phases for real fermions and the associated boundary excitations on the defects, which are classified by $\pi_{d-d_b-1}(\mathcal{M}^{s_{TC}}_{s_T s_C})$.  The configuration spaces for the fermionic excitations living on the defects are also given in Table \ref{tableM}, which turn out to be $d$-independent. Some of these localized excitations are the so-called Majorana fermions obeying non-abelian statistics \cite{MajoranaReturn}. Together with the two cases for the complex fermions as given in Table \ref{tableC}, in total we have ten classes for the fermionic topological phases.

Besides, we also have some cases for which only $C$ or $T$ symmetry exists. If there is only $C$ symmetry and $s_C=1$, then $\mathcal{M}_{s_C=1}=R^0_d$. On the other hand, for $s_C=-1$, $\mathcal{M}_{s_C=-1}=C_d$.

For the cases with only $T$ symmetry, it is not clear if there exists real representation for $\alpha^i$ and $\bm{M}$ since there is no $C$ symmetry. If there is no such representation but $U_t$ can be either real or pure imaginary, i.e., $U_t^2=\pm 1$, then $\mathcal{M}=C_{d+1}$. Otherwise, $\mathcal{M}_{s_T=1}=R^0_{d+1}$ and $\mathcal{M}_{s_T=-1}=R^1_d$, which are the same as the cases for $\mathcal{M}^+_{s_T+}$.

\section{SPT phases of real fermions in high energy physics}\label{sec4}

As we have seen there are eight real fermionic phases characterized by three parity parameters. Many realizations of them in the condensed matter systems are listed in Ref. \cite{Wen1}, and some  exhibit the interesting anyonic statistics \cite{MajoranaReturn}. On the other hand, in high energy physics one has more strict constraints. First, the high energy theory is continuum and the forms of $\alpha^i$ and $\bm{M}$ cannot be arbitrarily adjusted but mainly fixed by the Lorentz invariance. Second, the fermions considered here are fundamental particles, and cannot be treated as the composite one with non-trivial constituents. In contrast, the non-trivial fractionalization of the electrons are commonly postulated when considering the topological orders.

If we take the above considerations into account and also assume that the $C$ and $T$ symmetry operations will not involve the gauge and flavor physics, then it is easy to see that $s_C=1$. This is because the reality condition for fermion is given by
\be\label{realityc}
U_c \psi^*=\psi,
\ee
and perform once more the charge conjugation on \eq{realityc}, we will get $U_cU_c^* \psi=\psi$ which then implies $s_C=1$. Note that if the fermion is composite, then the reality condition may be defined up to some gauge transformations of the fractionalized charges, and the above argument could fail. From now on, we will assume $s_C=1$.

Since $U_c$ is unitary, then $U_cU_c^*=s_C=1$ which implies that $U_c$ is symmetric. For a (complex) symmetric matrix $A$, it is always possible to perform a Takagi's transformation such that $A=V D V^{\rm T}$ where $V$ is unitary and $D$ is a real nonnegative diagonal matrix whose entries are the nonnegative square roots of the eigenvalues of $A A^{\dagger}$ (Entry of ``Takagi's factorization" in  http://en.wikipedia.org/wiki/Matrix\_decomposition).  As $U_cU_c^{\dagger}=1$, one can bring $U_c$ to identity matrix after Takagi's transformation.

In fact, the Takagi's transformation on $U_c$ is a special case of the change of basis. This can be seen as follows. By keeping the relations \eq{CTsym-aM} form-invariant under the change of the basis, it is easy to see we must have
\be\label{TakagiT}
\tilde{U}_c=V U_c V^{\rm T}, \qquad  \tilde{U}_t=V^* U_t V^{\dagger} \qquad  \mbox{and}  \qquad  \tilde{\varGamma}^{\mu}=V \varGamma^{\mu} V^{\dagger},  \quad \mu=0,1,2,\cdots,d,
\ee
where $V$ is unitary. If $V$ is the Takagi's transformation, then $\tilde{U}_c=1$. In this basis, from Eqs. \eq{CTsym-a-C} and \eq{CTsym-M-C} we find that (we omit the tilde sign for simplicity in the following)
\be\label{realaM}
\alpha^{i*}=\alpha^i, \qquad \bm{M}^*=\bm{M}.
\ee
This implies that the Takagi transformation brings the Gamma matrices to the Majorana representation, i.e.,
\bes\label{Majoranarep}
\be\label{Majoranarep-1}
\varGamma^{\mu *}=\pm \varGamma^{\mu},
\ee
which also yields
\be\label{Majoranarep-2}
(m+\varLambda \varPhi)^*=\pm (m+\varLambda\varPhi).
\ee
\ees
Moreover, because $U_c$ is trivial, the Clifford algebra for K-theory analysis will be simply determined by $s_T$ if $U_t$ exists.   However, from Eq. \eq{G01-3} we find that $U_t^*=s_{TC} U_t$. It then seems that $s_{TC}$ will also play some role. This is actually not true because for $s_{TC}=-1$ it says that $U_t$ is pure imaginary. But we can compensate $U_t$ by an overall $i$ factor to make it real to form real Clifford algebra with $\alpha^i$ and $\bm{M}$.   This means that only $s_{TC}=1$ is allowed.  The configuration space is thus $\mathcal{M}^+_{s_T +}$ if $U_t$ exists.

        One can in fact relate $U_t$ to $U_c$ by comparing Eqs. \eq{CTsym-a-T} and \eq{CTsym-M-T} with Eqs. \eq{CTsym-a-C} and \eq{CTsym-M-C}. This leads to
\be\label{UtUc}
U_t=U_c^* \varGamma^0\varGamma,
\ee
up to an overall $i$ factor to make $U_t$ real, and
\be\label{gamma5}
\varGamma= (-1)^{(d-1)/4} \varGamma^0 \varGamma^1 \cdots \varGamma^d,
\ee
which obeys $\{ \varGamma,\varGamma^{\mu} \}=0$ and exists only for even space-time dimensions, i.e., $d+1=$even.  The pre-factor on the R.H.S. of Eq. \eq{gamma5} is chosen so that $\varGamma^2=1$.

 Using Eq. \eq{UtUc} and $s_C=1$ we find that
\be\label{sTdd}
s_T=U_t U_t^*= (-1)^{k_1+k_2},
\ee
where the exponents $k_{1,2}$ are defined by
\be\label{UcG0G5}
U_c \varGamma^{0*} U_c^{-1}=(-1)^{k_1} \varGamma^0,\qquad  U_c \varGamma^{*} U_c^{-1}=(-1)^{k_2} \varGamma.
\ee

It is also straightforward to see that $s_T$ (also $s_C$)  is preserved under Takagi's transformation \eq{TakagiT}.

To further proceed the classification, we need to determine if there exists Majorana representation \eq{Majoranarep-1} for the Gamma matrices so that Eq. \eq{realaM} holds. If so, there exists a $U_c$ with $U_cU_c^*=1$ and we can further determine $k_{1,2}$ and thus $s_T$ from Eqs. \eq{UcG0G5} and \eq{sTdd}.   The results depend on the space-time dimensions.   To check this,  we adopt some of the results in Ref. \cite{Polchinski}: In even space-time dimensions $d+1=2k+2$ \footnote{Our $d$ is different from the one in Ref. \cite{Polchinski} where $d=2k+2$.}, the matrices $\varGamma^{\mu *}$ and $-\varGamma^{\mu *}$ satisfy the same Clifford algebra as $\varGamma^{\mu}$, so that there exist unitary transformations $B_1$ and $B_2$ defined by
\bes\label{Bs}
\be\label{Bs-1}
B_1 \varGamma^{\mu *} B_1^{-1} = (-1)^k \varGamma^{\mu}, \qquad \mbox{and} \qquad  B_2 \varGamma^{\mu *} B_2^{-1} = (-1)^{k+1} \varGamma^{\mu}.
\ee
Moreover, these transformations also obey the following
\be\label{Bs-2}
B_1 \varGamma^* B_1^{-1} = B_2 \varGamma^* B_2^{-1}=(-1)^k \varGamma,
\ee
and
\be\label{Bs-3}
B_1B_1^*=(-1)^{k(k+1)/2}, \qquad \mbox{and} \qquad B_2 B_2^* =(-1)^{k(k-1)/2}.
\ee
\ees
Based on Eq. \eq{Bs} we can use $B_1$ or $B_2$ to construct $U_c$ which obeys $U_cU_c^*=1$ and yields Majorana representation \eq{Majoranarep-1} after Takagi's transformation.

If $U_c$ does not involve gauge and flavor physics, we can identify $U_c$ as either $B_1$ or $B_2$ and then determine $k_{1,2}$ and thus $s_T$. From Eq. \eq{Bs-3} we can find that the Majorana representation exists only for $d=0,1,2,3,7$ mod $8$.  To be more specific,  we can conclude that
\begin{itemize}

\item for $d=0,2 \mod 8$,  there exists $U_c$ such that $U_c=1$ after Takagi's transformation,  and there exists no $U_t$ to transform all the Gamma matrices uniformly so that $U_t$ plays no role in the Clifford algebra. The configuration space for the topological phases of the real fermions is $\mathcal{M}=R_{d \mod 8}^0$. Moreover, for $d=0 \mod 8$, $m+\varLambda\varPhi$ is pure imaginary; and for $d=2 \mod 8$, $m+\varLambda\varPhi$ is real. However, $d=0$ is exceptional because $m+\varLambda\varPhi$ can be either real or pure imaginary.

\item for $d=1,3,7 \mod 8$, we have $U_c=1$ and $U_t= \varGamma^0 \varGamma$ after Takagi's transformation. By identifying $U_c$ and determining $s_T$,  we find that the configuration space for the topological phases of the real fermions for $d=1$, $\mathcal{M}^+_{++}=R^0_2$ for real  $m+\varLambda\varPhi$ and $\mathcal{M}^+_{-+}=R^0_0$ for pure imaginary $m+\varLambda\varPhi$; for $d=3$, $\mathcal{M}^+_{-+}=R^0_2$ for real  $m+\varLambda\varPhi$; and for $d=7$, $\mathcal{M}^+_{++}=R^0_0$ for pure imaginary  $m+\varLambda\varPhi$.

\item for $d=5,6 \mod 8$, there is no Majorana fermion.
\end{itemize}

We can turn on the scalar condensate $\varPhi$, then $U_c$ could involve the physics in the gauge sector.  If $\varPhi$ is real, then it plays no role in $U_c$ and does not affect the reality condition of $\bm{M}$, also the condition in determining $s_T$. Thus, the above results for $\varPhi=0$ also work for the cases with $\varPhi\ne 0$ but real.

On the other hand, $\varPhi$ could be pseudo-real, i.e., exists a unitary transformation $G$ (in the gauge/color space) such that
\be
G \varPhi^* G^{-1}=-\varPhi,
\ee
with $GG^*=-1$. For example, $\varPhi$ is in the pseudo-real representation of $SU(2)$, i.e., $\varPhi=\varPhi^a \tau^a$ and $G={\rm i}\tau^2$ where $\tau^a$'s are the Pauli matrices. This naively leads to the obstruction for realizing the reality condition of $\bm{M}$.  We can, however, bypass this no go by introducing the charge conjugation in the following way:
\be
U_c=B\otimes G,
\ee
where $B$ is either $B_1$ or $B_2$ in Eq. \eq{Bs}, which obeys $BB^*=-1$. Since the Gamma matrices do not involve the gauge and flavor sectors, they transform under $U_c$ in the same way as in Eqs. \eq{Bs-1} and \eq{Bs-2}, i.e., $U_c \varGamma^{\mu *} U_c^{-1}=\pm \varGamma^{\mu}$. More important is
\be
U_c (\varGamma^0\otimes  \varPhi)^* U_c^{-1}=\mp \varGamma^0\otimes  \varPhi,
\ee
so that $U_c \bm{M}^* U_c^{-1}=\bm{M}$ for
\be
m^*=\pm m \qquad \mbox{and}  \qquad \varLambda^*=\mp \varLambda,
\ee
which is a more detailed version of Eq. \eq{Majoranarep-2} when $\varPhi$ is pseudo-real.

Note that $U_c$ must be symmetric so that we can perform Takagi's transformation to bring it to unity matrix and fulfill the reality condition \eq{realaM}.   Thus we need to look for $B$ such that $BB^*=-1$, that is, the pseudo-real representation of Gamma matrices before Takagi's transformation.  From Eq. \eq{Bs-3} we find such representations exist if $d=3,4,5,6,7$ mod $8$.  After identifying $B$ and determining $s_T$, we can conclude that
\begin{itemize}
\item for $d=4,6 \mod 8$, there exists $B$ with $BB^*=-1$ such that $U_c=1$ after Takagi's transformation, but no $U_t$ for enlarging the Clifford algebra. The configuration space for the topological phases of the real fermions is $\mathcal{M}=R^0_{d \mod 8}$. Moreover, for $d=4$, $m+\varLambda\varPhi$ should be pure imaginary and for $d=6$ it should be real.

\item for $d=3,5,7 \mod 8$, there exists $B$ with $BB^*=-1$ such that $U_c=1$ and $U_t=\varGamma^0 \varGamma$ after Takagi's transformation.  We then find that the configuration space for the topological phases of the real fermions for $d=3$, $\mathcal{M}^+_{++}=R^0_4$ for pure imaginary $m+\varLambda\varPhi$; for $d=5$, $\mathcal{M}^+_{++}=R^0_6$ for real $m+\varLambda\varPhi$ and $\mathcal{M}^+_{-+}=R^0_4$ for pure imaginary $m+\varLambda\varPhi$; and for $d=7$, $\mathcal{M}^+_{-+}=R^0_6$ for real  $m+\varLambda\varPhi$.

\item for $d=0,1,2 \mod 8$, there exists no pseudo-real fermion. But $d=0$ is exceptional because $\varGamma^0={\rm i}\sigma^1$ with $B={\rm i}\sigma^2$ can fulfill the pseudo-real condition.

\end{itemize}

We now summarize the above results for both real and pseudo-real $\varPhi$ and
the associated classifications of the edge modes on the defects for QFT$_{d+1}$ in Table \ref{tableHEP}.

\begin{table}% [htdp]
\caption{Classification of symmetry-protected topological order of real free
fermions in $(d+1)$-dimensional Minkowski space.   Here $\mathcal{M}_{d_b}$ denotes
the configuration space for the topological phases on the $(d_b+1)$-dimensional defect,
 and the edge modes are then classified by $\pi_0(\mathcal{M}_{d_b})$. The second row indicates
 the scalar condensate $\varPhi$ is  ``real"(R) or ``pseudo-real"(PR).  The third row indicates the
 Dirac gamma matrices are real or purely imaginary. The table is mod $8$ implied by Bott's periodicity.
 However, $d=0$ is exceptional with possible real configurations of R(Im), R(Re) (omitted) and PR(Im)
 (omitted), but all classified by $\Pi_0(R^0_0)=\mathcal{Z}_2$. See the discussions in the main text for details.}
\begin{center}
\begin{tabular}{ c c c c c c }
\hline
$d\mod 8$ \quad & R or PR \quad &$\varGamma^{\mu}$ \quad &$\mathcal{M}$ \quad&$\pi_0(\mathcal{M})$ \quad&$\mathcal{M}_{d_b}$ \\ \hline
$d=0$ & R  &Im & $R_0^0$ &$\bm{Z}_2$ &$R_{d_b+1}^0$ \\
$d=1$ & R  &Re & $R^0_2$ &$\bm{Z}$ & $R^0_{d_b+2}$ \\
      & R  &Im & $R^0_0$ &  $\bm{Z}_2 $ &$R^0_{d_b}$ \\
$d=2$ & R  &Re & $R^0_2$ &$\bm{Z}$ &  $R_{d_b+1}^0$ \\
$d=3$ & R  &Re &$R^0_2$ & $\bm{Z}$ & $R^0_{d_b}$ \\
      & PR &Im & $R^0_4$ & $0$ &$R^0_{d_b+2}$ \\
$d=4$ & PR & Im &$R^0_4$& $0$ &  $R^0_{d_b+1}$ \\
$d=5$ & PR &Re & $R^0_6$ & $\bm{Z}$ & $R^0_{d_b+2}$ \\
      & PR &Im & $R^0_4$ & $0$ & $R^0_{d_b}$ \\
$d=6$ & PR &Re &$R^0_6$& $\bm{Z}$ & $R^0_{d_b+1}$ \\
$d=7$ & R  &Im &$R^0_0$& $\bm{Z}_2$ &$R_{d_b+2}^0$ \\
      & PR & Re&$R^0_6$&$\bm{Z}$&$R_{d_b}^0$\\ \hline
\end{tabular}
\end{center}
\label{tableHEP}
\end{table}%

We then wonder if there are possible topological insulators/superconductors realized in the models of particle physics. In reality, most of the particles in standard model and beyond are charged and interact via strong, electromagnetic and weak forces, thus the above classification scheme for free fermions do not apply. However, the neutrinos are the exceptional because they only interact via weak force, which is weak enough so that the neutrinos can be thought as the free fermions. Moreover, the neurtinos's mass structure is not fully understood  and one can imagine it may arise from some see-saw mechanism involving more heavy weakly interacting partners.  Taking into account of these unseen heavy partners (spectator flavors) with the already-known three flavors of standard model neutrinos to have large number of flavors, the system just fit to the setup for the above K-theory analysis for the symmetry-protected topological orders. Especially, the neutrinos are by themselves the Majorana fermions, then the possible topological phases depend only on time reversal symmetry since the $C$ and $P$ symmetries break badly in standard model. If we neglect the small $CP$ (or equivalently $T$ by $CPT$ theorem) violation, then according to $d=3$ entry in Table \ref{tableHEP} the topological fermionic phases are classified by $\pi_0(R^0_2)=\bm{Z}$ for real $\varPhi$ and $\pi_0(R^0_4)=0$ for pseudo-real $\varPhi$.  Otherwise, it will be classified by $\pi_0(R_3^0)=0$. This exemplifies how the symmetry-violating perturbations will affect the possible topological phases for the neutrinos.

One may also extend the above consideration into the brane-world models, then the standard model particles such as neutrinos live on the $(3+1)$-dimensional branes embedded in the bulk space-time. That is, the brane-world can be viewed as a $d_b=3$ defect in the $(d+1)$-dimensional space-time. The result will then depend on the dimension $d$ of the bulk space-time. If we assume no $CP$ violation, then according to Table \ref{tableHEP} the topological phases are classified by $\pi_0(R^0_3)=0$ for $d=5,7$, by $\pi_0(R^0_4)=0$ for $d=4,6,8$, and by $\pi_0(R^0_5)=0$ for $d=5,7,9$. There is simply no nontrivial real topological phases associated with neutrino sector for the brane-world models if there is no $CP$ violation. It is interesting to see here how the nature of Clifford algebras constrains the topological phases of the neutrinos in the brane-world models. One can perform the similar analysis for more general scenarios.

\section{Classification of holographic fermionic SPT phases}\label{sec5}

We now would like to classify the patterns of fermionic operator mixing (or the SPT phases) in the dual CFTs. As discussed, the task is equivalent to classifying the topological classes of the free massive  fermions in the co-dimensional one higher Minkwoski space.  Naively it seems that we can just carry the result of Table \ref{tableHEP} directly into the classification for the dual CFTs.   It turns out this is to be case for the classification of the complex fermions, i.e., without imposing $C$ and $T$ symmetries, but not the case for the classification of the real fermions. Below we will elaborate the subtlety.

When considering the dual mapping between the bulk fermions and the boundary fermionic operators in the AdS/CFT correspondence,  there is an immediate issue to be taken care: the number of components of the irreducible representations for the bulk and the one for the boundary fermions are different.  More concretely, for even $d$, the boundary fermionic operators are in the Weyl representation, whose number of components is half of the one of the bulk Dirac fermions; and for odd $d$, the number of components of the boundary Dirac fermionic operators is also only half of the one of the bulk Dirac fermions.

Thus, we need to project half of the bulk fermion's degrees of freedom to match the boundary ones. Moreover, the form of the projector is suggested by the boundary action, which is required to fulfill the variational principle in deriving equation of motion and impose the Dirichelt boundary condition. The boundary action (in the context of AdS/CFT correspondence but not for the bulk/edge correspondence of the classification scheme)  takes the form
\be\label{bndyGr}
\pm {\rm i} \int {\rm d}^dx \sqrt{-g_b } \bar{\psi} \varGamma^r \psi,
\ee
where $\varGamma^r$  is the Gamma matrix along the radial direction of AdS space. Thus, we can choose $\varGamma^r$ as the projector and decompose $\psi$ into
\be\label{psipm}
\psi=\psi_+ +\psi_-, \qquad \mbox{with}  \qquad \varGamma^r \psi_{\pm}=\pm \psi_{\pm}.
\ee
Then, Eq. \eq{bndyGr} is reduced to
\be \label{bndyaction}
\pm \; {\rm i} \int {\rm d}^dx\sqrt{-\gamma} \; \bar{\psi}\psi= \pm \; {\rm i} \int {\rm d}^dx\sqrt{-\gamma} \; (\bar{\psi}_+ \psi_-+\bar{\psi}_-\psi_+).
\ee
The plus (minus) sign corresponds to project out $\psi_-$ ($\psi_+$). These two choices are called the standard and alternative quantization schemes \cite{Klebanov:1999tb,Iqbal:2009fd}, respectively \footnote{There are also the other boundary actions such as the ones considered in Ref. \cite{Laia:2011zn} for holographic flat band. Different boundary actions yield different boundary CFTs which are related by the RG flows caused by the double trace operators \cite{Laia:2011zn,Laia:2011wf}. }.

With the boundary action \eq{bndyaction} (by picking up either sign of $\pm$) we can then impose the Dirichlet boundary condition on either one of  $\psi_{\pm}$, which can then be identified as source coupled to its dual boundary fermionic operator.  If we are considering the complex fermions,  this prescription by picking only one of $\psi_{\pm}$ as the dual of the CFT fermionic operator will then match the degrees of freedom of bulk fermion and its CFT dual without further complication. Thus,   the topological phases of the complex fermions for the dual CFTs are classified by $\pi_0(C_d)$.

On the other hand, when considering the real fermions, to fulfill the above prescription we also shall require $\psi_{\pm}$ to be real.  Once we pick up the Majorana representation \eq{Majoranarep-1} after appropriate Takagi's transformation, this is equivalent to requiring $\varGamma^{\mu}$ (including $\varGamma^r$) to be real, not pure imaginary.

\begin{table}%[htdp]
\caption{Classification of the SPT phases of real fermion in holographic CFTs. The letters R and PR denote as ``real" $\varPhi$ and ``pseudo-real" $\varPhi$, respectively. We restrict ourselves to $d\le 6$ since AdS space with dimension higher than this may not be stable due to the large back reaction from the perturbations.
}
\begin{center}
\begin{tabular}{ c c c c c  }
\hline
AdS$_{d+1}$/CFT$_d$ &$(s_C, s_T, s_{TC})$&$\varPhi$ &$\mathcal{M}$ &$\pi_0(\mathcal{M})$ \\\hline
 CFT$_1$ &  $(+,+,+)$ & R&  $R_2^0$ & $\bm{Z}$ \\
 CFT$_2$ &     $(+,\times, \times )$ & R & $R_2^0$ & $\bm{Z}$  \\
 CFT$_3$ & $(+,-,+)$ &  R & $R_2^0$ & $\bm{Z}$  \\
 CFT$_4$ &  $( \times ,-, \times )$ &R/PR  &  $C_5$ & $0$     \\
 CFT$_5$ &   $(+,+ , +)$ &PR & $R_6^0$ & $\bm{Z}$  \\
 CFT$_6$ & $(+,\times , \times)$ &PR & $R_6^0$& $\bm{Z}$\\
\hline
\end{tabular}
\end{center}
\label{tableF}
\end{table}%

Therefore, to classify the topological phases of the real fermions for the dual CFTs, we need not only the Majorana representations but also the ones with real Gamma matrices. After examining the relations \eq{Bs} (the results are summarized in Table \ref{tableHEP}) we find that this kind of representations exist for $d=1,2,3$ with real $\varPhi$, and for $d=5,6$ with pseudo-real $\varPhi$. For $d=4$ there is no such representation but $U_t$ exists so that the topological phases for this case is classified by $\pi_0(C_5)=0$. We summarize the above results  in the Table \ref{tableF}. Note that these classifications yield the possible patterns for the fermionic operator mixings in the dual CFTs.

It is interesting to see that all the topological phases of the real fermions in AdS spaces considered above are classified by $\bm{Z}$ except AdS$_5$ as shown in Table \ref{tableF}.

\section{Conclusion and Discussion}\label{sec6}

In this paper, we adopt the K-theoretical classification of TI/TSc to the multi-flavor free fermions in the context of high energy physics as well as AdS/CFT correspondence and the results are summarized in Table \ref{tableHEP} and  \ref{tableF}. The latter corresponds to classify the possible patterns of fermionic operator mixing in the dual CFTs, and the results could be useful for the study of level crossings when tuning the relevant parameters beyond the large $N$ limit.   Our scheme can also be seen as a trial to classify the fermionic topological phases for the holographic CFTs, though it is still a challenging problem to define or even classify the topological phases of the gapless systems. Based on our results, it is interesting to further study the dynamical implication of these topological edge modes and how they can be probed holographically. Especially, we would hope to make connections of these topological edge modes with anyonic statistics, as usually expected for the Majorana fermions.

\begin{table}%[htdp]
\caption{Classification of SPT phases of real fermions in holographic IR gapped QFT$_{d-1}$ dual to the $(d+1)$-dimensional AdS soliton.}
\begin{center}
\begin{tabular}{ccc}
\hline
AdS$_{d+1}$/QFT$_{d-1}$ & ~$\mathcal{M}$ ~& ~$\pi_1(\mathcal{M})$ \\\hline
QFT$_1$ &  $R_2^0$ & $\bm{Z}_2$  \\
QFT$_2$ &  $R_2^0$ & $\bm{Z}_2$  \\
QFT$_3$ &   $C_5$ & $\bm{Z}$     \\
QFT$_4$ &   $R_6^0$ &$0$  \\
QFT$_5$  & $R_6^0$ &  $0$ \\
\hline
\end{tabular}
\end{center}
\label{tableS}
\end{table}%

In this paper, we only consider the pure AdS space, which is dual to the CFTs. We may also extend our consideration to other asymptotically AdS backgrounds. One interesting example is the AdS solitons which are believed to be dual to some gapped systems. In this case, the gap is introduced in a geometric way by smoothly capping off the IR geometry. The edge modes may then be localized at the IR endpoint, which is holographically dual to the IR effective dual theory living on a co-dimensional two space-time. To classify the topological orders associated with these edge modes, one can invoke the bulk/edge correspondence, which then yields $\pi_1(\mathcal{M})$. The result is briefly summarized in Table \ref{tableS}. However, in the literatures there is almost no study about the solutions of the Dirac equation in this kind of backgrounds so that the full holographic dictionary is not clear. Further study for this case will be illuminating.

\section*{Acknowledgements}
  We thank Xiao-Gang Wen for the discussion on the issue of K-theory analysis of models in high energy physics. FLL is supported by Taiwan's Ministry of Science and Technology (MoST) grants (100-2811-M-003-011 and 100-2918-I-003-008). We thank the support of National Center for Theoretical Sciences (NCTS).

\section*{Conflict of interest}
The authors declare that they have no conflict of interest.

\end{document}